\documentclass[letterpaper,10pt,twocolumn]{article}
\usepackage{ol2}

\usepackage{epsfig,amsmath,graphicx,amssymb}
\def\be{\begin{equation}}
\def\ee{\end{equation}}
\def\bee{\begin{eqnarray}}
\def\ene{\end{eqnarray}}
\def\bes{\begin{subequations}}
\def\ees{\end{subequations}}

\begin{document}

\twocolumn[
\title{Linear superpositions of gap solitons in periodic Kerr media}

\author{Yuliy V. Bludov$^1$,  Vladimir V. Konotop$^2$, and Mario Salerno$^3$}

\address{ $^1$ Centro de F\'{\i}sica, Universidade do Minho,
Campus de Gualtar, Braga 4710-057, Portugal
\\
$^2$ Centro de F\'{\i}sica Te\'orica e Computacional and Departamento de F\'{\i}sica, Faculdade de Ci\^encias, Universidade de Lisboa,
Avenida Professor Gama
Pinto 2, Lisboa 1649-003, Portugal
\\
 $^3$ Dipartimento di Fisica ``E.R. Caianiello'', CNISM  and INFN - Gruppo Collegato di Salerno, Universit\`a di Salerno, Via Ponte don Melillo, 84084 Fisciano (SA), Italy
}


\begin{abstract}
The existence of a novel type of solitons in periodic  Kerr media  constructed  as superposition of  noninteracting gap-solitons of different kinds (bright, dark and periodic) is first demonstrated. The  periodic modulation of the nonlinearity is used to suppress the cross phase modulation between components to allow  the superimposed beam to propagate for long distances.
\end{abstract}
]

\maketitle

\noindent

Photonic structures with periodic modulations of   parameters are ideal systems  to investigate   nonlinear properties of wave dynamics (see e.g.~\cite{KartVyslTorn}).
In this context, arrays of optical waveguides  are one of the most used configurations for practical
implementations of various physical phenomena involving optical solitons. To  avoid problems of cross-phase modulations affecting  beams consisting of different modes, which  differ by frequencies and/or by wavelengths, the nonlinear properties of the  light  propagation in these structures  have been  investigated mainly for monochromatic light. Instability induced by cross-phase modulation, however, could arise also for monochromatic light, if linear superposition of nonlinear modes of different wavelengths are considered.
Meantime, stabilizing effects on the dynamics achieved by means of periodic modulations of the nonlinearity have been explored in the context of  Bose-Einstein condensates in optical lattices where it has been shown that periodic modulations of the interatomic interaction allow one to induce Bloch oscillations~\cite{SKB08}, intra-band  (Rabi) oscillations~\cite{SKB09_2} and dynamical localization~\cite{SKB09} of gap solitons which persist for long times (long propagation distances in the optical context). In view of the analogies between photonic lattices  and Bose-Einstein condensates in optical lattices it is of interest to investigate the possibility of using periodic modulations of the nonlinearity as a tool for controlling the cross phase modulation in superimposed beams.

The aim of this Letter is to demonstrate the existence of a novel type of solitons in periodic  Kerr media  made as superpositions of  noninteracting gap-solitons of different wavevectors and different kinds (bright, dark and periodic). The  periodic modulation of the nonlinearity is used to suppress the cross phase modulation between components to allow  the superimposed beam to propagate for long distances.

As model equation for the  propagation of a monochromatic beam  along $z$-axis in a periodic Kerr media we use the following nonlinear Schr\"odinger equation for the dimensionless electric field envelope  $q$ (see e.g.~\cite{KartVyslTorn}):
\begin{eqnarray}
\label{eq:NLS}
iq_z=-q_{xx}+V(x)q+G(x) |q|^2q.
\end{eqnarray}
Here $V(x)=V(x+\pi)$ and  $G(x)=G(x+\pi)$ denoting the linear refractive index and the nonlinearity modulations in the transverse direction, respectively. Hereafter we fix  $V(x)=V\cos(2x)$ and $G(x)=\sigma+G\cos(2x)$, both taken of period $\pi$ without loss of generality.  The     average   Kerr index $\sigma$, as well as amplitudes of linear refractive index, $V$,  and of nonlinearity, $G$, modulations are considered as free parameters, used below for achieving zero cross phase modulation conditions.
\begin{figure}[htb]
\begin{center}
\vspace{-0.2cm}
\includegraphics[width=\columnwidth]{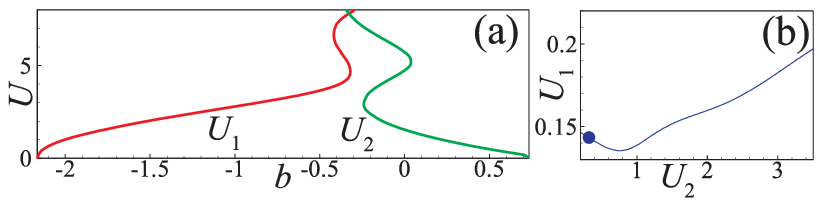}
\includegraphics[width=\columnwidth]{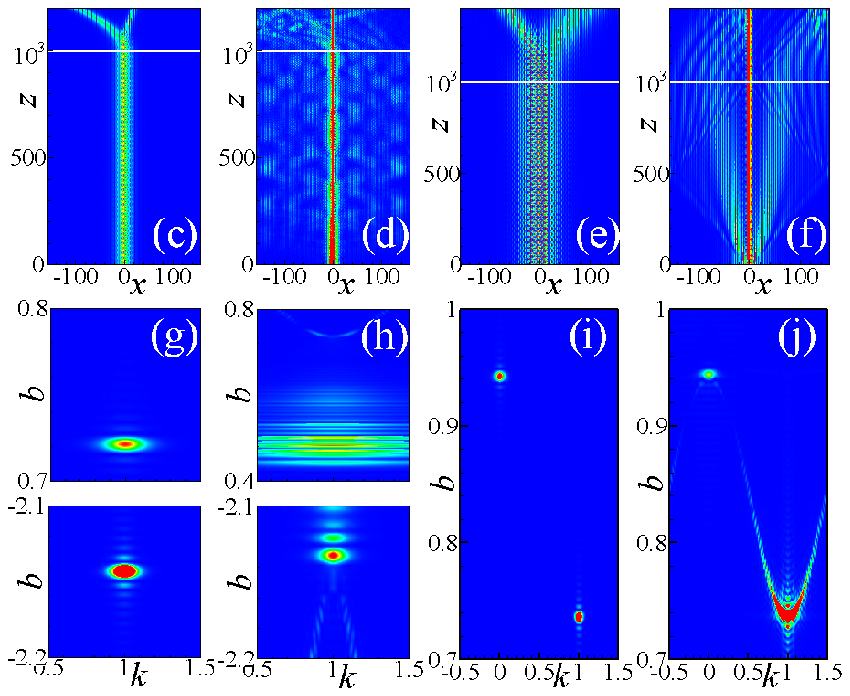}
\end{center}
\vspace{0.2cm}
\caption{ (color on line) (a)
Families of the  modes bifurcating from the lower, $U_{1}$, and
upper $U_{2}$ edges  of the first gap: $b\in[-2.1659,0.7332]$ (the
edges coincide with the panel boundaries);  (b)
$U_{1}$ {\it vs} $U_{2}$ for the solitons fulfilling
(\ref{eq:cross-zero}); (c--f) Intensity distributions $|q(x,z)|^2$ and (g--j) spectra $|Q(b,k)|^2$ at distance $z_0=1000$ of input  beams (\ref{eq:lin-comb}) consisting of two superimposed  solitons with parameters: (c,g)  $U_{1}=0.32$ ($b_1=-2.15$),  $U_{2}=0.143$ ($b_2=0.71$); (d,h)  $U_{1}=0.645$ ($b_1=-2.1$), $U_{2}=0.33$ ($b_2=0.6$); (e,i)
$U_{1}=2.77$ ($b_1=0.732$),  $U_{2}=1.7$ ($b_2=0.938$); (f,j) $U_{1}=2.77$ ($b_1=0.732$), $U_{2}=5.28$ ($b_2=1.0$).
In panels (c,d,g,h) solitons are near edges of the first gap while in panels (e,f,i,j) near edges of the first band. Other parameters are $\sigma=-1$,  $V=-3$,  $G=3.05$ (a--d,g,h), $G=1.282$ (e,f,i,j).
}
\vspace{-0.5cm}
\label{fig:soliton-dz}
\end{figure}
The periodicity of the linear refractive index induces a band structure determined by the eigenvalue problem $ d^2\varphi_{\alpha,k} /dx^2-V(x)\varphi_{\alpha,k} =b_\alpha(k)\varphi_{\alpha,k}$,
where the
eigenvalues  $b_\alpha(k)$ (propagation constants) are periodic functions of the Bloch vector $k\in[-1,1]$  and the associated orthonormal set of eigenmodes $\varphi_{\alpha,k}$ satisfy the Bloch condition $\varphi_{\alpha,k}(x + \pi) = e^{i k \pi} \varphi_{\alpha,k}(x)$ (here $\alpha$ denotes the
band index). We are particularly interested in the modes bordering different edges of the same or of different bands.
Respectively, we use subscripts $l$, $u$  to denote lower and upper propagation constant values ($b_l<b_u$).

The self-phase modulation of   small amplitude solitons bifurcating from band edges  $b_{l,u}$ is described by $\chi_ {l,u} =\int_{0}^{\pi} G(x) |\varphi_{l,u}|^4 dx$, while the cross-phase modulation is determined by $\chi   =\int_{0}^{\pi} G(x) |\varphi_{l}|^2|\varphi_{u}|^2dx$. Then  necessary conditions for the existence of the bright solitons, combined with the requirement for zero cross-phase modulation, allowing   them to bifurcate independently from different edges, take the form
\begin{eqnarray}
\label{eq:cond}
D_{l} \chi_{l}<0,  \qquad D_{u} \chi_{u}<0, \qquad \chi=0,
\end{eqnarray}
where $D_{l,u}$ denote  the linear diffraction coefficients of the respective $l,u$ modes: $D_{l,u}=- d^2b_{l,u}/dk^2$ (see \cite{SKB08} for details).
These conditions can be satisfied for proper choices of $V(x)$ and $G(x)$.

Let us now consider a superposition of two localized solutions, $w_{1,2}(x)e^{ib_{1,2}z}$ of Eq.~(\ref{eq:NLS}), i.e.
\begin{eqnarray}
q(x,z)=w_1(x)e^{ib_1z}+w_2(x)e^{ib_2z+i\Theta}.
\label{eq:lin-comb}
\end{eqnarray}
Here without loss of generality the functions $w_j(x)$ ($j=1,2$) are taken to be real solutions of the stationary equation $d^2w_j/dx^2+b_jw_j  -V(x)w_j+G(x)w_j^3=0$ while the constant $\Theta$ denotes a relative phase. For propagation constants sufficiently close to the gap edges, i.e. for $b_1 \approx b_l$ and $b_1 \approx b_u$, these solutions can be represented as $w_{1,2}(x)=A_{1,2}(\mu_{1,2}x)\varphi_{l,u}(x)$, where $\mu_{1,2}=\sqrt{|b_{1,2}-b_{l,u}|}\ll 1$ is a small parameter characterizing detunings to the gap, and $A_j(\cdot)$ are slowly varying amplitudes. Then (\ref{eq:cond}) ensures that $\chi\sim \mu_j^2\ll \chi_{l,u}$ so that  the cross phase modulation of Bloch states  implies  strong  suppression of the cross-phase modulation of the small amplitude solitons $w_{1,2}(x)$. Respectively, we expect the superposition of such solutions  to approach a stable solutions of the nonlinear equation. When the detuning towards the gap increases,  the condition (\ref{eq:cond}) must be modified, to ensure
\begin{equation}
\int_{-\infty}^\infty w_{1}w_{2}dx=0,\quad \mbox{and}\quad
\int_{-\infty}^\infty G(x)w^2_1 w^2_2 dx=0.
\label{eq:cross-zero}
\end{equation}

To show that when (\ref{eq:cross-zero}) are satisfied the superposition  (\ref{eq:lin-comb}) becomes highly accurate approximation of the exact  solutions
we recall that Eq.~(\ref{eq:NLS}) admits  two conserved quantities: the energy flow $U[q]=\int_{-\infty}^\infty|q|^2dx$ and the Hamiltonian $H[q]=\int_{-\infty}^\infty\left[|  q_x |^2+V(x)|q|^2+\frac 12 G(x)|q|^4 \right]dx$. For the solutions (\ref{eq:cross-zero}), then the above conditions imply:
$U[q]=U[w_1]+U[w_2]$ and $H[q]=H[w_1]+H[w_2]$ so that, in spite of the nonlinearity in Eq.~(\ref{eq:NLS}), the superimposed beam (\ref{eq:lin-comb}) will propagate as an exact solution, without appreciable distortions. In the case of  solitons with $b_{1,2}$ bifurcating from opposite edges of the first gap [Fig.~\ref{fig:soliton-dz}(a)], the first of conditions (\ref{eq:cross-zero}) is satisfied due to the parity  of solutions $w_1(x)$ (odd) and $w_2(x)$ (even). As to the cross-phase modulation,
for fixed values of $V$ and $G$, for each even soliton with $b_2$ (and the energy flux $U_2=U[w_2]$) there exists its unique odd counterpart with $b_1$ (and the energy flux $U_1=U[w_1]$),  satisfying the second condition in (\ref{eq:cross-zero}).
We remark that for two low-amplitude solitons the second condition in (\ref{eq:cross-zero}) is usually satisfied for  $G \gtrsim G_0$, where $G_0$ denotes the amplitude of $G(x)$ for which  $\chi=0$ (for the refractive index parameters of Fig.\ref{fig:soliton-dz}(a) $G_0\approx 2.98$ while $G=3.05$).
For this situation the respective dependence $U_1(U_2)$ is illustrated in Fig.~\ref{fig:soliton-dz}(b). We however emphasize that only stable points of the presented curve can be used in an experiment. Typically, only small-amplitude solitons are stable:  $U_{1}\lesssim 0.5$ for the parameters chosen in Fig.~\ref{fig:soliton-dz}.

The fact that the superposition can propagates without significant distortion is confirmed by Figs.~\ref{fig:soliton-dz}(c,e). In order to emphasize the distortionless propagation of the beam, at $z_0=1000$ (denoted in Figs.~\ref{fig:soliton-dz}(c-f) by horizontal lines) we applied a refractive index gradient in the transverse direction, modeled by adding the term $0.001xq(x,z)$ in the right-hand side of Eq.~(\ref{eq:NLS}) for $z>z_0$.  We observe  that  when conditions (\ref{eq:cross-zero}) are met [Figs.~\ref{fig:soliton-dz}(c,e)] the two superimposed solitons propagate undistorted until the refractive gradient is applied and after that it splits into two gap solitons propagating in opposite directions, due to the opposite signs of the diffraction coefficients $D_{l,u}$. If, however, Eq. (\ref{eq:cross-zero}) is strongly violated we get that the superimposed beam is  destroyed after relatively short propagation distances (i.e. for $z\ll z_0$) [see Figs.~\ref{fig:soliton-dz}(d,f)].
\begin{figure}
\begin{center}
\includegraphics*[width=\columnwidth]{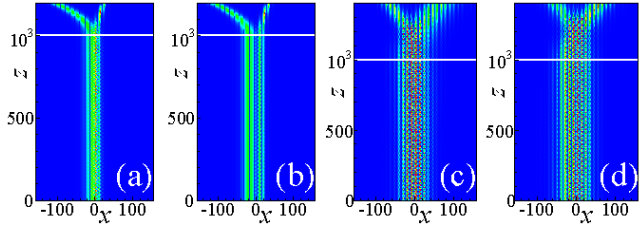}
\end{center}
\caption{ (color on line)
Propagation of two superimposed gap solitons with centers
shifted  at the input by  $\Delta x=4\pi$ (panels a,c), or by $\Delta x=10\pi$ (panels b,d).
Soliton parameters are the same as in Fig.~\ref{fig:soliton-dz}(c) (panels a,b), or Fig.\ref{fig:soliton-dz}(e) (panels c,d).
}
\label{fig:soliton-shifted}
\end{figure}
The spectra $Q(b,k)=\int_{-\infty}^{\infty}   q(x,z)e^{-ibz} \varphi_{\alpha,k}(x)dzdx$ of the superimposed beams computed at $z=z_0$ are shown in Figs.~\ref{fig:soliton-dz}(g-j) (each panel corresponding to the spatial dynamics shown in the middle row just above it). We see that when the zero cross phase modulation condition is satisfied [Figs.~\ref{fig:soliton-dz}(g,i)]  the output spectrum is centered around the same parameters   $b,k$ in the Fourier space as the respective input beam (in spite of the very long propagation distance), while in the opposite case, [Figs.~\ref{fig:soliton-dz}(h,j)], the spectrum becomes strongly deformed. Note the differences between  spectral widths and transverse spatial extension of the superimposed beam in Figs.~\ref{fig:soliton-dz}(c,g) and the ones in Figs.~\ref{fig:soliton-dz}(e,i). While for the former case (component solitons at opposite edges of the first gap) the orthogonality  condition is always satisfied due to the parity, in the latter case (component solitons at opposite edges of the first band) this is true only in the small amplitude limit, as we remarked before  (in this case the orthogonality of the solitons can be achieved only through the orthogonality of the  Bloch states).
This explains the wider width in the transverse direction of the latter beam  and the narrowing of the spectral widths of the component gaps-solitons.

In Fig.~\ref{fig:soliton-shifted} we have shown the effect of an initial  displacement $\Delta x$  on the propagation of the two superimposed solitons in Fig.~\ref{fig:soliton-dz}(c,e). We see that the superpositions remain stable both for small
[Figs.\ref{fig:soliton-shifted}(a,c)] and large [Figs.\ref{fig:soliton-shifted}(b,d)] displacements. This is understood by the fact that, being the solitons of  small amplitudes, nonlinear-orthogonality condition (\ref{eq:cross-zero}) is only slightly violated by small displacements.

\begin{figure}
\begin{center}
\includegraphics*[width=\columnwidth]{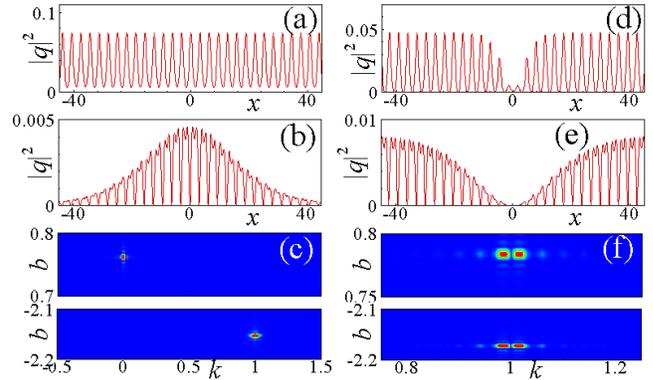}
\end{center}
\caption{ (color on line) (a,b) Even periodic background with the energy flux per period $U_{2}=\int_{0}^{\pi}w_2^2(x)dx=0.103$ and $b_2=0.76$ (panel a),  superimposed with an odd soliton having $U_{1}=0.132$ and $b_1=-2.16$ (panel b) for the parameters $\sigma=-1$, $V=-3$,  $G=6$;  (d,e) Superposition of two dark solitons bordering edges of the  gap $b\in[-2.1659,0.7332]$: even dark soliton with   $b_2=0.78$ (panel d) and odd dark soliton with   $b_1=-2.18$ (panel e) for $V=3$, $\sigma=1$, $G=2.95$; (c,f) Spectrum $|Q(b,k)|^2$ relative to superpositions of the waves  in (a,b) and in (d,e) respectively, both taken at
$z_0=1000$.
}
\label{fig:soliton-pw-dark}
\end{figure}

The above results can be extended to superpositions
of nonlinear periodic waves with a bright soliton [Fig.~\ref{fig:soliton-pw-dark}(a--c)] and to superpositions of dark solitons [Fig.~\ref{fig:soliton-pw-dark}(d--f)].
In the first case the periodic (even) wave $w_2(x)$  bifurcates from the lower edge of semi-infinite gap [Fig.~\ref{fig:soliton-pw-dark}(a)], while   the bright (odd) soliton $w_1(x)$ bifurcates from the lower edge of first gap [Fig.~\ref{fig:soliton-pw-dark}(b)]. The both entities are stable; we however note the difference in their amplitudes [c.f. the scales of $|q|^2$ in Figs.~\ref{fig:soliton-pw-dark}(a) and \ref{fig:soliton-pw-dark}(b)]. In the second example, the dark solitons bifurcate from the opposite  edges of the first gap towards the first [Fig.~\ref{fig:soliton-pw-dark}(d)] and the second [Fig.~\ref{fig:soliton-pw-dark}(e)] allowed bands. In both cases the condition for zero cross phase modulation is the same as in (\ref{eq:cross-zero}) (or $\chi=0$ in the small amplitude limit) but  conditions for  existence and stability of the modes must be changed as $D_{l} \chi_{l}<0$,  $\;  D_{u} \chi_{u}>0$ and $\; D_{l,u} \chi_{l,u}>0$ for periodic wave-bright soliton and dark soliton - dark soliton superpositions, respectively.  In the bottom panels of  Fig.~\ref{fig:soliton-pw-dark} are shown the output spectra computed at $z=1000$ of the corresponding superimposed modes. In both cases we see that the linear superposition propagates in stable manner for a long distance as consequence of  the zero cross phase modulation between  beam components. Similar results can be derived for solitons subjected to refractive indexes and nonlinearity modulations of different periods, as the ones considered in Ref.~\cite{SakMalom}.

In conclusion, we have demonstrated  the possibility of stable propagation of new types of superimposed beams in periodic Kerr media  which are constructed as linear  superpositions of gap solitons of several types and different wavelengths. The Kerr nonlinearity was designed  in such a manner that the cross-phase modulation between beam components is eliminated or strongly reduced.
These results can be  of practical interest for multi-component parallel transmission and manipulation of optical signals in nonlinear media.

\smallskip

VVK was
supported by FCT (Portugal) under the grant PTDC/FIS/112624/2009.
 MS acknowledges support from MIUR through  a PRIN-2008 initiative.

\pagebreak

1. Y. V. Kartashov, B. A. Malomed, and  L. Torner, "Solitons in Nonlinear Lattices," Rev. Mod. Phys. {\bf 83}, 247-305 (2011)

2. M. Salerno, V. V. Konotop, and Yu. V. Bludov, "Long-Living Bloch Oscillations of Matter Waves in Periodic Potentials,"  Phys. Rev. Lett. {\bf 101}, {030405} (2008).

3. Yu.V. Bludov, V. V. Konotop,  and M. Salerno, "Rabi oscillations of matter-wave solitons in optical lattices," Phys. Rev. A {\bf 80}, {023623} (2009).

4. Yu. V. Bludov, V. V. Konotop,   and M. Salerno, "Dynamical localization of gap-solitons by time periodic forces," Europhys. Lett. {\bf 87}, {20004} (2009).

5. H. Sakaguchi  and B. A. Malomed, "Solitons in combined linear and nonlinear lattice potentials" Phys. Rev. A {\bf 81}, 013624 (2010)


\end{document}